\documentclass[12pt,preprint]{aastex}

\newcommand{\Ha}{\mbox{H${\alpha}$}}
\begin{document}
\title{Panoramic GALEX FUV and NUV imaging of M31 and M33}
\author{
David A. Thilker\altaffilmark{1},
Charles G. Hoopes\altaffilmark{1},
Luciana Bianchi\altaffilmark{1}, 
Samuel Boissier\altaffilmark{2},
R. Michael Rich\altaffilmark{3},
Mark Seibert\altaffilmark{4},
Peter G. Friedman\altaffilmark{4},
Soo-Chang Rey\altaffilmark{4,5},
Veronique Buat\altaffilmark{6},
Tom A. Barlow\altaffilmark{4},
Yong-Ik Byun\altaffilmark{5},
Jose Donas\altaffilmark{6},
Karl Forster\altaffilmark{4},
Timothy M. Heckman\altaffilmark{7},
Patrick N. Jelinsky\altaffilmark{8},
Young-Wook Lee\altaffilmark{5},
Barry F. Madore\altaffilmark{2,9},
Roger F. Malina\altaffilmark{6},
Christopher Martin\altaffilmark{4},
Bruno Milliard\altaffilmark{6},
Patrick F. Morrissey\altaffilmark{4},
Susan G. Neff\altaffilmark{10},
David Schiminovich\altaffilmark{4},
Oswald H. W. Siegmund\altaffilmark{8},
Todd Small\altaffilmark{4},
Alex S. Szalay\altaffilmark{7},
Barry Y. Welsh\altaffilmark{8}, and
Ted K. Wyder\altaffilmark{4}}

\altaffiltext{1}{Center for Astrophysical Sciences, The Johns Hopkins
University, 3400 N. Charles St., Baltimore, MD 21218, dthilker,choopes,bianchi@pha.jhu.edu}

\altaffiltext{2}{Observatories of the Carnegie Institution of Washington,
813 Santa Barbara St., Pasadena, CA 91101, boissier@ociw.edu, barry@ipac.caltech.edu}

\altaffiltext{3}{Department of Physics and Astronomy, University of
California, Los Angeles, CA 90095, rmr@astro.ucla.edu}

\altaffiltext{4}{California Institute of Technology, MC 405-47, 1200 East
California Boulevard, Pasadena, CA 91125, mseibert,friedman,screy,krl,cmartin,patrick,ds,tas,wyder@srl.caltech.edu, tab@ipac.caltech.edu}

\altaffiltext{5}{Center for Space Astrophysics, Yonsei University, Seoul
120-749, Korea, byun@obs.yonsei.ac.kr, ywlee@csa.yonsei.ac.kr}

\altaffiltext{6}{Laboratoire d'Astrophysique de Marseille, BP 8, Traverse
du Siphon, 13376 Marseille Cedex 12, France, veronique.buat,jose.donas,roger.malina,bruno.milliard@oamp.fr}

\altaffiltext{7}{Department of Physics and Astronomy, The Johns Hopkins
University, Homewood Campus, Baltimore, MD 21218, heckman,szalay@pha.jhu.edu}

\altaffiltext{8}{Space Sciences Laboratory, University of California at
Berkeley, 601 Campbell Hall, Berkeley, CA 94720, patj,ossy,bwelsh@ssl.berkeley.edu}

\altaffiltext{9}{NASA/IPAC Extragalactic Database, California Inst.
of Tech., Mail Code 100-22, 770 S. Wilson Ave., Pasadena, CA 91125}

\altaffiltext{10}{Laboratory for Astronomy and Solar Physics, NASA GSFC, Greenbelt, MD 20771, neff@stars.gsfc.nasa.gov}

\begin{abstract}
We present {\it Galaxy Evolution Explorer} (GALEX) far-UV and near-UV
mosaic observations covering the entirety of M31 and M33.  For both
targets, we measure the decline of surface brightness (in FUV and NUV)
and changes in FUV--NUV color as a function of galactocentric radius.
These UV radial profiles are compared to the distribution of ionized
gas traced by $\Ha$ emission.  We find that the extent of the UV emission,
in both targets, is greater than the extent of the observed \ion{H}{2}
regions and diffuse ionized gas.  We determine the ultraviolet diffuse
fraction in M33 using our FUV observations and compare it to the
H$\alpha$ diffuse fraction obtained from wide-field narrow-band
imaging.  The FUV diffuse fraction appears to be remarkably constant
near 0.65 over a large range in galactocentric radius, with departures
to higher values in circumnuclear regions and, most notably, at the
limit of the $\Ha$ disk.  We suggest that the increase in FUV diffuse
fraction at large galactocentric radii could indicate that a
substantial portion of the diffuse emission beyond this point is not
generated in situ but rather scattered from dust, after originating in
the vicinity of the disk's outermost \ion{H}{2} regions.  Radial
variation of the $\Ha$ diffuse fraction was also measured.  We found
the $\Ha$ diffuse fraction generally near 0.4 but rising toward the
galaxy center, up to 0.6.  We made no attempt to correct our diffuse
fraction measurements for position-dependent extinction, so the quoted
values are best interpreted as upper limits given the plausibly higher
extinction for stellar clusters relative to their surroundings.

\end{abstract}







\section{Introduction}
\label{sintro}

A fundamental goal of the {\it Galaxy Evolution Explorer} (GALEX)
mission is to gauge the history of star formation in the Universe over
the interval $0 < z < 2$.  This (forthcoming) analysis will depend on
integrated FUV and NUV flux measurements for $>10^7$ distant galaxies,
anticipated upon completion of the GALEX imaging surveys.  However,
accurate interpretation of these data requires a comprehensive
understanding of the possibly heterogeneous resolved UV source
population amongst galaxies of diverse type.  Also, detailed
rest-frame UV observations of nearby galaxies are critically needed as
a benchmark for optical observations of galaxy morphology at high
redshift.  For these reasons, in addition to the intrinsic interest in
bright, well-resolved targets, the GALEX Nearby Galaxy Survey (NGS:
Bianchi et al. 2004a, b) is being undertaken.

We have obtained FUV and NUV GALEX NGS observations of M31 and M33,
covering the full extent of each galaxy.  Local Group objects
constitute an especially critical portion of the NGS sample, as our
angular resolution affords separation of individual young stellar
clusters from their neighbors and host environment.  Furthermore,
these nearby targets provide an excellent opportunity to measure
properties of the apparently diffuse UV emission, uncontaminated by
embedded stellar clusters which ultimately generate the extended
radiation field (although some ``diffuse'' emission may also originate
from faint, unresolved stars).

In this Letter, we describe preliminary analysis of the GALEX M31 and
M33 mosaics.  We characterize the overall UV morphology and contrast
the GALEX sources with corresponding ionized nebulae
(\ion{H}{2} regions, shells, filaments, and diffuse gas).  For
reference, Keel (2000) provides a succinct review of previous UV
investigations in M31 and M33.

\section{Observations and Data Analysis}

We analyzed FUV (1350--1750~\AA) and NUV (1750--2750~\AA) imaging data
from GALEX, plus pre-existing ground-based $\Ha$ observations. For
Local Group targets within the NGS, the GALEX mission plan calls for
two integrations (conducted during separate orbits), amounting to a
total of $\sim$ 3 ks per field.  The imagery presented here represents
most of the eventual M31/M33 database, but one pointing in M33 and a
few in M31 remain to be observed for the second time.  We note that
fields probing some portions of Andromeda's extreme outer HI disk and
high-velocity cloud population (Thilker et al. 2004, Braun et al. in
prep) are not yet available.

The tiling of M31, which incorporates some coverage gaps, was dictated
by the need to avoid bright stars.  Fifteen $1.25\arcdeg$ diameter
fields were observed. Central coordinates and total exposure times are
given in Table 1.  M33 has been surveyed in its entirety without any
gaps, using one ``first-look'' exposure of the northern disk and a
7-pointing hexagonal mosaic centered on the galaxy.  The majority of
M33 lies within the central pointing of this mosaic.  Figures 1 and 2
present our GALEX (FUV, NUV) color-composite images of M31 and M33.


{\footnotesize
\begin{deluxetable}{lccclcccc}
\tablewidth{0pt}
\tablecaption{GALEX FIELDS FOR M31 AND M33 {\label{tlimits}} }
\tablehead{
\multicolumn{1}{l}{Field}         & \multicolumn{1}{c}{$\alpha_{J2000}$}    & \multicolumn{1}{c}{$\delta_{J2000}$} &
\multicolumn{1}{c}{Exp. (s)}
}
\startdata
M31-F1&00:42:37.01&41:13:22.08&650\\
M31-F3&00:45:00.00&41:51:24.84&1704\\
M31-MOS0&00:40:40.11&40:49:57.25&2437\\
M31-MOS1&00:35:30.52&41:05:53.34&2180\\
M31-MOS2&00:39:24.06&41:51:18.50&1637\\
M31-MOS3&00:45:34.56&40:32:05.78&2374\\
M31-MOS4&00:44:46.35&41:32:38.22&2750\\
M31-MOS5&00:33:25.82&39:57:06.66&1642\\
M31-MOS6&00:45:11.80&39:34:21.00&3075\\
M31-MOS7&00:43:33.38&42:34:07.90&2762\\
M31-MOS8&00:48:40.83&42:01:58.15&2699\\
M31-MOS9&00:36:26.40&42:45:00.72&3265\\
M31-MOS10&00:29:42.56&40:44:45.92&1452\\
M31-MOS11&00:48:51.16&42:57:45.14&1879\\
M31-MOS12&00:41:23.84&40:21:34.52&3031\\
\hline
M33&01:33:52.27&31:08:19.61&1660\\
M33-MOS0&01:33:50.90&30:39:36.00&3399\\
M33-MOS1&01:31:53.70&31:22:50.02&3409\\
M33-MOS2&01:29:58.40&30:39:22.00&3366\\
M33-MOS3&01:31:55.60&29:56:13.99&1959\\
M33-MOS4&01:35:46.30&29:56:15.00&3269\\
M33-MOS5&01:37:43.40&30:39:24.01&3035\\
M33-MOS6&01:35:48.00&31:22:50.99&3171
\enddata
\end{deluxetable}
}
\normalsize



For M31, we compared the GALEX observations with $\Ha$ data from 10
KPNO-4m/MOS fields belonging to the NOAO ``Survey of Local Group
Galaxies Currently Forming Stars'' (Massey et al. 2001).  In M33, we
used a Burrell Schmidt $\Ha$ mosaic (Hoopes \& Walterbos 2000)
covering all of the disk.

We adopt distances of 770 and 840 kpc, respectively, to M31 and M33
(M31: Freedman \& Madore 1990, M33: Freedman, Wilson, \& Madore 1991).
The $5\arcsec$ ~FWHM GALEX PSF corresponds to $\sim20$ pc.  RMS flux
sensitivity varies with position. Typical 1$\sigma$ limits are 6.6 and
$2.8\times10^{-19}$ erg s$^{-1}$ cm$^{-2}$ \AA$^{-1}$ for FUV and NUV,
respectively, evaluated at the scale of the PSF.  Expressed in terms
of surface brightness, these limits correspond to 28.0(28.2) AB mag
arcsec$^{-2}$ in FUV(NUV).  This sensitivity is sufficent to detect
substantial diffuse emission throughout both galaxies.

\section{UV morphology and correlation with $\Ha$}
\label{sfits}

Figure 3 presents FUV, NUV, and FUV--NUV radial profiles for both
galaxies.  For M33, we also show a radial profile for diffuse $\Ha$
emission.  We have measured the median (sky-background-subtracted)
surface brightness in concentric elliptical annuli to $2\arcdeg$ along
the major axis for M31 and $0.6\arcdeg$ for M33, stepping by
15\arcsec.  A complete description of the procedure is given by
Bianchi et al. (this volume). In M31, to the limit of our sensitivity,
there is no well-defined ``edge'' of the stellar disk traced by the
GALEX observations.  At least out a galactocentric distance of 27 kpc,
there is no significant downturn in the surface density of
intermediate mass stars producing the observed UV radiation in M31.
This is interesting given the much more obvious limit to the extent of
\ion{H}{2} regions traced in $\Ha$ imaging ($\sim 20$ kpc).
Enhancements in the median surface brightness occur at radii dominated
by Andromeda's well-known 10 and 15 kpc arms. In M33, at the disk's
edge, we note that the UV emission drops off significantly slower than
the median $\Ha$ emission.  Furthermore, we see no evidence for a UV
edge at 32' reported by Buat (1994), most likely a consequence of our
improved sensitivity.  On both FUV--NUV panels, we indicate
Bruzual \& Charlot (2003) model colors for starbursts spanning a range in
age, and three model colors appropriate for differing periods of continuous
star formation.  The plotted models are for solar metallicity.
When interpreting the measured colors in relation to the starburst
models, one may consider the age corresponding to the best matching
model color as the age of the last major star formation episode.  In
both galaxies, the FUV--NUV color of the typical stellar population
generally becomes bluer with increasing radius, however the magnitude
of this effect is variable.  In M33, the outward trend toward bluer UV
color is slight and FUV--NUV remains remarkably constant over the
majority of the galactocentric radii examined.  This reflects the
global extent of recent star formation in M33.  Almost every locale
within M33 has undergone a significant burst within the past $\sim
200$ Myr.  See Bianchi et al. (this volume) for similar analysis of
M51 and M101.

The proximity of our Local Group targets provides an excellent
opportunity for decomposition of the GALEX imaging data into discrete
(compact) sources and diffuse emission, especially so for M33 due to
its favorable inclination ($i = 57\arcdeg$).  As in the study of diffuse
ionized gas (DIG, Hoopes et al. 2001), the UV diffuse fraction
(defined as the integral of diffuse emission / integral of all
emission) can be used to consider issues of radiation transport in the
ISM.  The UV diffuse fraction acts as a indicator for the origin and
subsequent path of non-ionizing photons, whereas the $\Ha$ diffuse
fraction provides an indirect tracer of the Lyman continuum.

We measured the diffuse fraction (in FUV, NUV, and $\Ha$) as a
function of radius in M33 using our GALEX data and the Schmidt mosaic
of Greenawalt (1998).  This was accomplished by computing the average
ratio of two images in elliptical annuli oriented to match the M33
disk.  The first of these two images was a representation of the
diffuse emission on 1.5 kpc scales, generated by applying a large
circular median-filter to the data.  The median operator effectively
eliminated the flux from all discrete structures smaller than 1.5 kpc.
The second image contributing to our measured ratio was a smoothed
version of the original data, generated by convolving with a Gaussian
kernel sized to achieve resolution (1.5 kpc) matching the median image
described above.  Both these images were background subtracted before
computation of the diffuse fraction.  To verify the accuracy of our
method in the UV, we compared with published work on the diffuse
ionized gas. Consistent with Hoopes \& Walterbos (2000), we find that
the $\Ha$ diffuse fraction is approximately 0.4 at radii hosting the
bulk of M33's emission line gas.  We determined the UV diffuse
fraction in M33 using both FUV and NUV images, but found the NUV
estimate is notably contaminated (in spots) by foreground stars.
Accordingly, we adopt the FUV-band for gauging the UV diffuse
fraction.

Figure 4 presents the UV and $\Ha$ diffuse fractions measured for M33.
The ultraviolet diffuse fraction appears to be remarkably constant
near 0.65 over a large range in galactocentric radius, with departures
to higher values in circumnuclear regions and, most notably, at the
limit of the $\Ha$ disk.  The increase in FUV diffuse fraction at
large galactocentric radii may reflect that a substantial portion of
the diffuse emission beyond this point is not generated in situ but
rather scattered from dust, after originating in the vicinity of the
disk's outermost \ion{H}{2} regions.  Note that we have not attempted
correction of the UV diffuse fraction for decreased extinction in the
field away from compact star forming regions, but it is possible that
the UV diffuse fraction could adjust substantially downward once this
is considered (Buat et al. 1994).  Radial variation of the $\Ha$
diffuse fraction was also measured.  We found values generally near
0.4 but rising toward the galaxy center, up to 0.6.

The relative intensity of diffuse UV and $\Ha$ emission and UV color
in DIG as a function of distance from associated stellar clusters is
another useful probe of the ISM.  Indeed, the ionization mechanism of
the DIG has long been a puzzle, as it requires a very large (30-50$\%$
of the ionizing photons from a galaxy's OB stellar pop.), sustained
energy source for the gas to remain ionized.  If OB stars in cluster
environments are responsible for the ionization, the Lyman continuum
photons must ``leak'' from \ion{H}{2} regions.  Hoopes \& Walterbos
(2000) first considered the contribution of field OB stars as an
additional way to support the DIG, finding that they can account for
roughly 40\% of the observed emission.  Our GALEX observations will
soon be used to measure extinction corrected FUV/H$\alpha$ and FUV--NUV
within the DIG, comparing FIR, NIR bands with our FUV image to gauge
extinction (Thilker et al. in prep.).

The GALEX FUV and NUV imagery of M31 and M33 allows for a detailed
study of the relation between resolved emission line structures
(traced by $\Ha$) and young stellar clusters (and/or single massive
stars).  For this Letter, we present an overview of the issue. Figure
5 shows a color-composite image of $\Ha$ + continuum (red), continuum
(green), and GALEX NUV (blue) imagery for the whole of M33, whereas
Fig. 6 presents an analogous high-resolution view for a spiral arm
segment in M31.  [Note that the full-quality figures of this paper are
electronically available, and retain much more detail than in print.]
The GALEX imagery exhibits nearly one to one correlation with the
distribution of \ion{H}{2} regions (as expected), but notably also
shows UV bright clusters or star inside the vast majority of apparent
$\Ha$ bubbles.  Although $\Ha$ is a very sensitive tracer of massive
star formation, it reveals only those sources which are still very
young.  GALEX provides a more complete census of recent star formation
as it traces young-to-moderate age stellar populations, up to several
hundred Myr, unless they are highly obscured.  This conclusion is
obvious from inspection of Figs. 5 and 6, and serves to highlight the
most basic premise of the GALEX mission: to survey cosmic star
formation history in the vacuum ultraviolet.

\begin{acknowledgments}
GALEX (Galaxy Evolution Explorer) is a NASA Small
Explorer, launched in April 2003. We gratefully acknowledge NASA's
support for construction, operation, and science analysis for the
GALEX mission.
\end{acknowledgments}



\clearpage

\begin{figure}
\epsscale{0.85}
\plotone{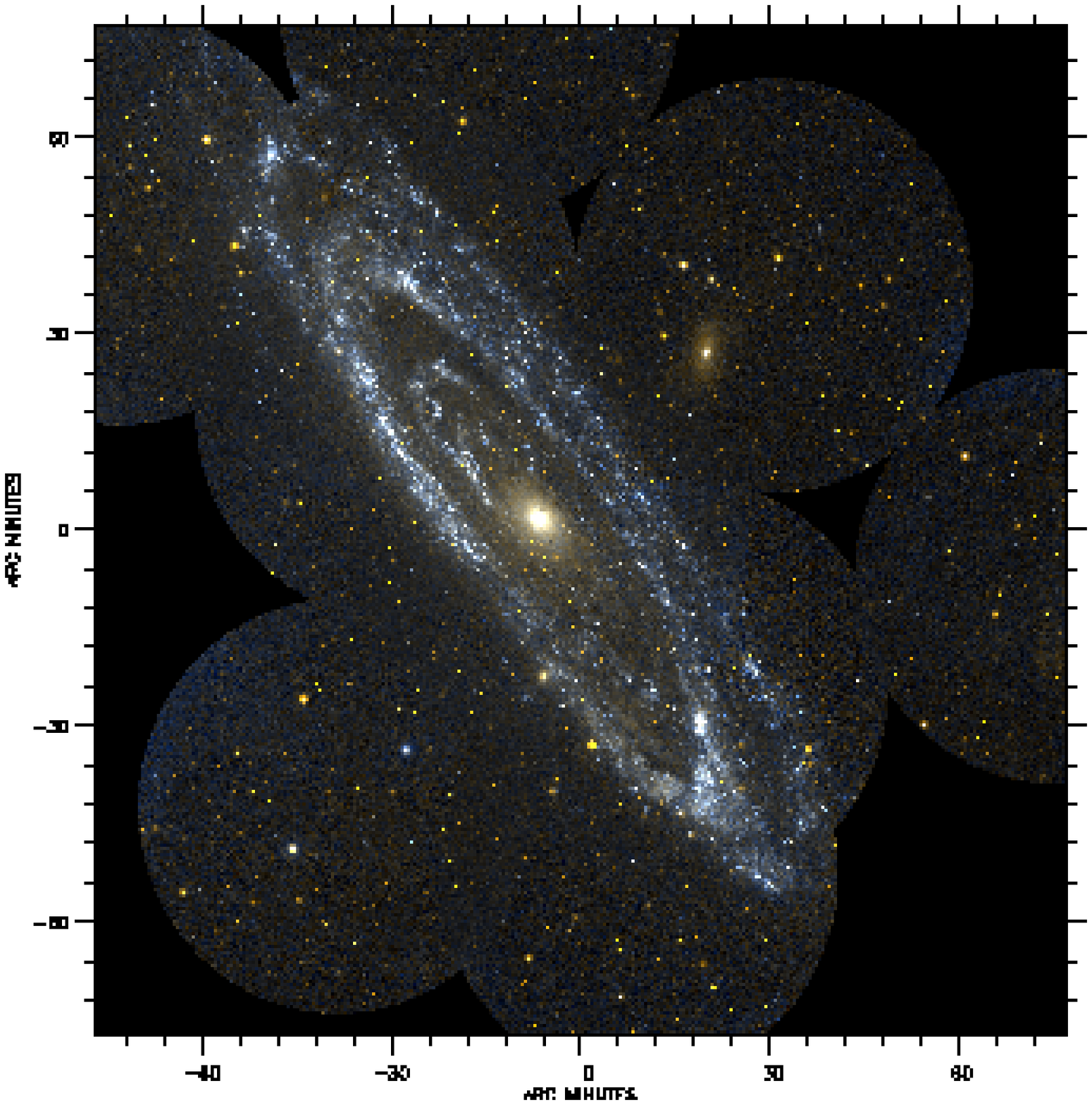}
\caption{\small The centermost $3\times3\arcdeg$ region of the GALEX M31 mosaic.
The image, a composite representation of our FUV and NUV data, has
been created so as to make sources brightest in FUV appear blue and
those features brightest in NUV appear orange.  Regions which are
especially bright in both bands are rendered nearly white.  Note that
two of the early ``test'' pointings listed in Table 1 (M31-F1, M31-F3)
and depicted here lie within the bright disk of the galaxy, where they
were reimaged with the subsequent grid of ``mosaic'' fields.  Six of the 15 M31 fields do
not appear in this image (MOS1,5,6,9,10,11) as they are located
outside the bounds of the figure.  The conspicuous gap in coverage to
the SW of Andromeda's southern tip is caused by the need to avoid a
bright field star.  Later in the GALEX mission, the bright star limit will be relaxed. When this takes place, we will observe the M31 coverage gaps as much as practical. \label{foverlay} }
\end{figure}

\begin{figure}
\epsscale{0.85}
\plotone{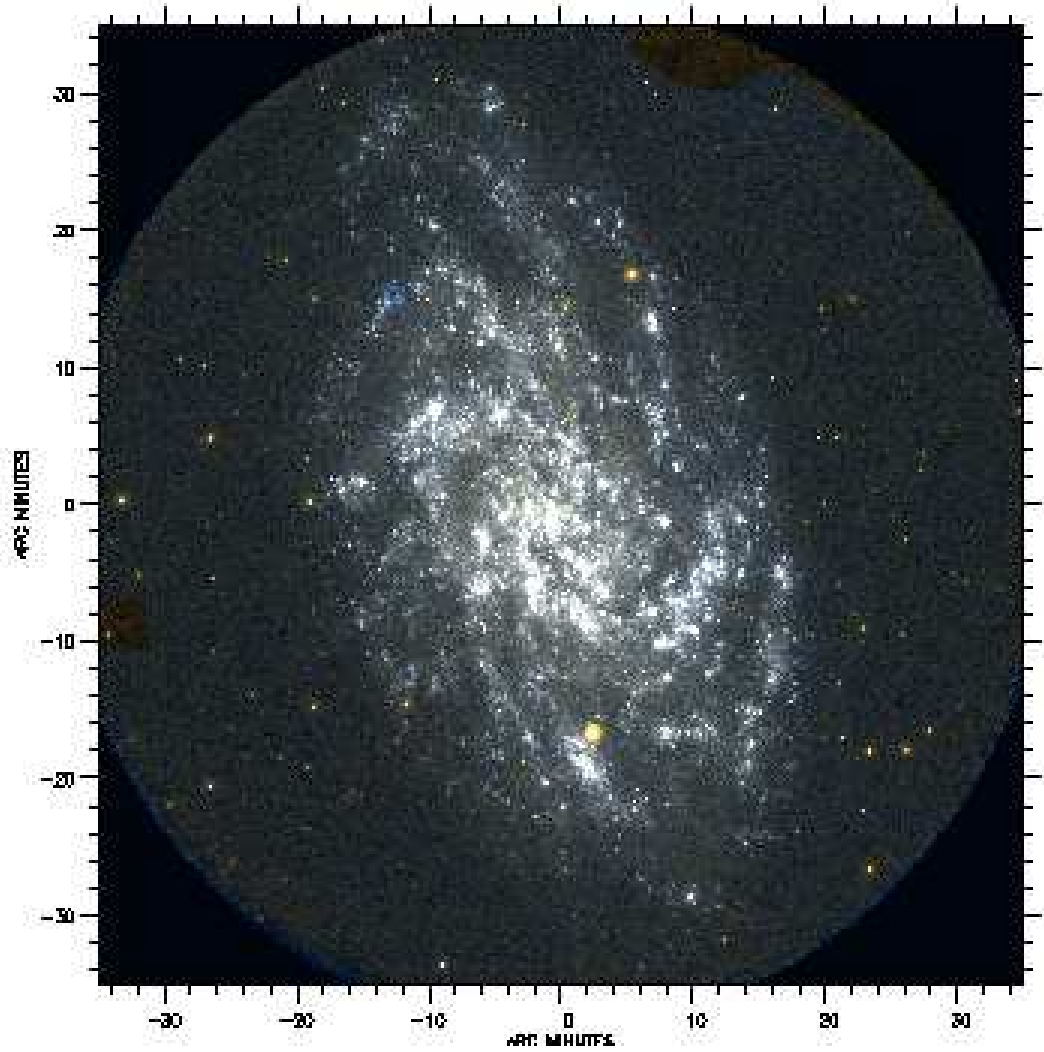}
\caption{The central field of the M33 GALEX mosaic.  The
color-composite image has been created as for M31 in Fig. 1, with
FUV-bright regions appearing blue.  A contiguous area extending
$\sim$ $2.6\arcdeg$ in diameter around M33 has already been observed.}
\end{figure}

\begin{figure}
\epsscale{1.}
\plotone{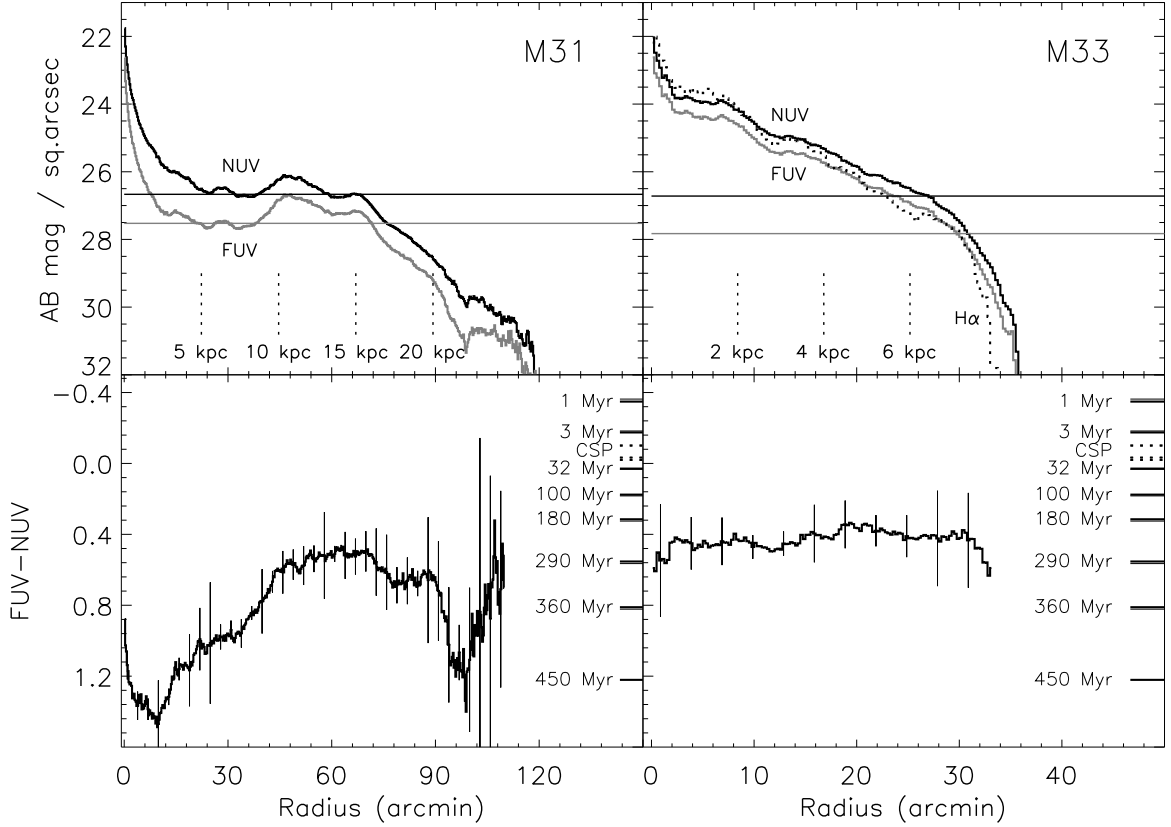}
\caption{Radial profiles of FUV, NUV surface brightness and FUV--NUV
color in M31 and M33.  On the top panels of the figure we plot the
median surface brightness (AB mag / sq. arcsec) in both bands.  The
dark line indicates NUV, whereas the light colored line shows FUV.  We
also plot the estimated sky background with horizontal lines.  An
$\Ha$ profile for M33 is also plotted, with the normalization adjusted
to match the FUV at 30\arcmin ~radius.  Note the remarkable range in
surface brightness probed from the inner disk to the outer disk.  For
this measurement, we took the median value within each concentric
elliptical annulus so as to gauge the ``background'' underlying
stellar population, rather than the (luminosity-weighted) mean which
is biased toward the clusters which are presently UV-brightest.
The bottom panels show median FUV--NUV with error bars ($\pm 1\sigma$)
indicated on the profile.  We also plot intrinsic and reddened
synthetic colors for instantaneous burst populations of varied age and
three prolonged periods of continuous star formation (CSP: 100 Myr, 1
Gyr, 10 Gyr).  In M31, note the strong FUV--NUV indication for rather
minimal recent SF (during the past 500 Myr) within the interarm gap
between Andromeda's bulge and 5 kpc arm.  Our measurements have not
been corrected for the variable extinction which is bound to exist.}
\end{figure}

\begin{figure}
\epsscale{0.85}
\plotone{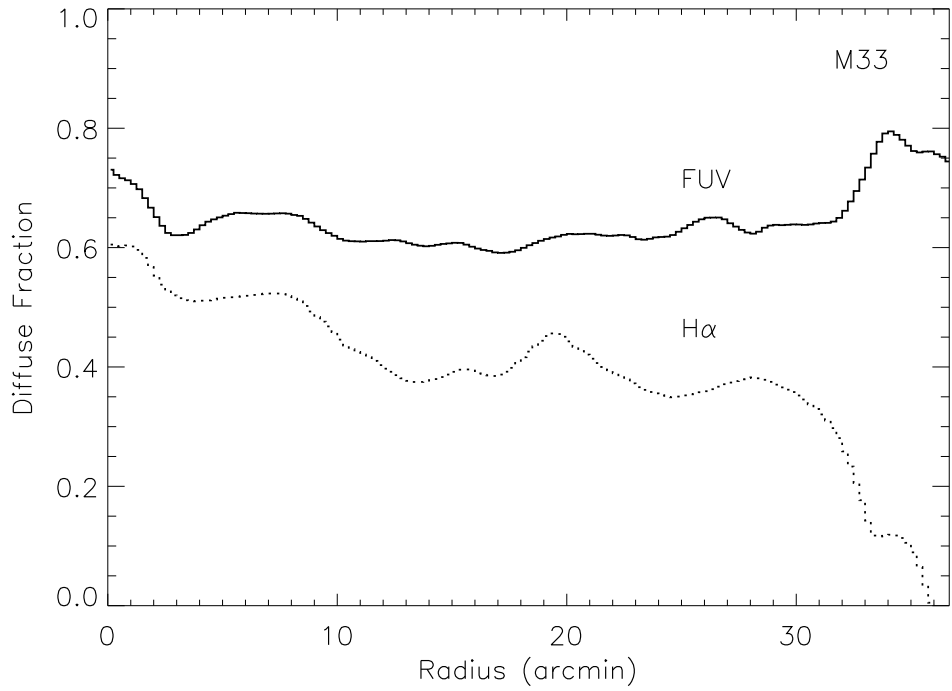}
\caption{The diffuse fraction in FUV and $\Ha$ as a function of
galactocentric radius in M33.  The effective resolution of these
curves is $\sim 400 \arcsec$, owing to the method used to isolate the
DIG from compact sources.}
\end{figure}

\begin{figure}
\epsscale{0.85}
\plotone{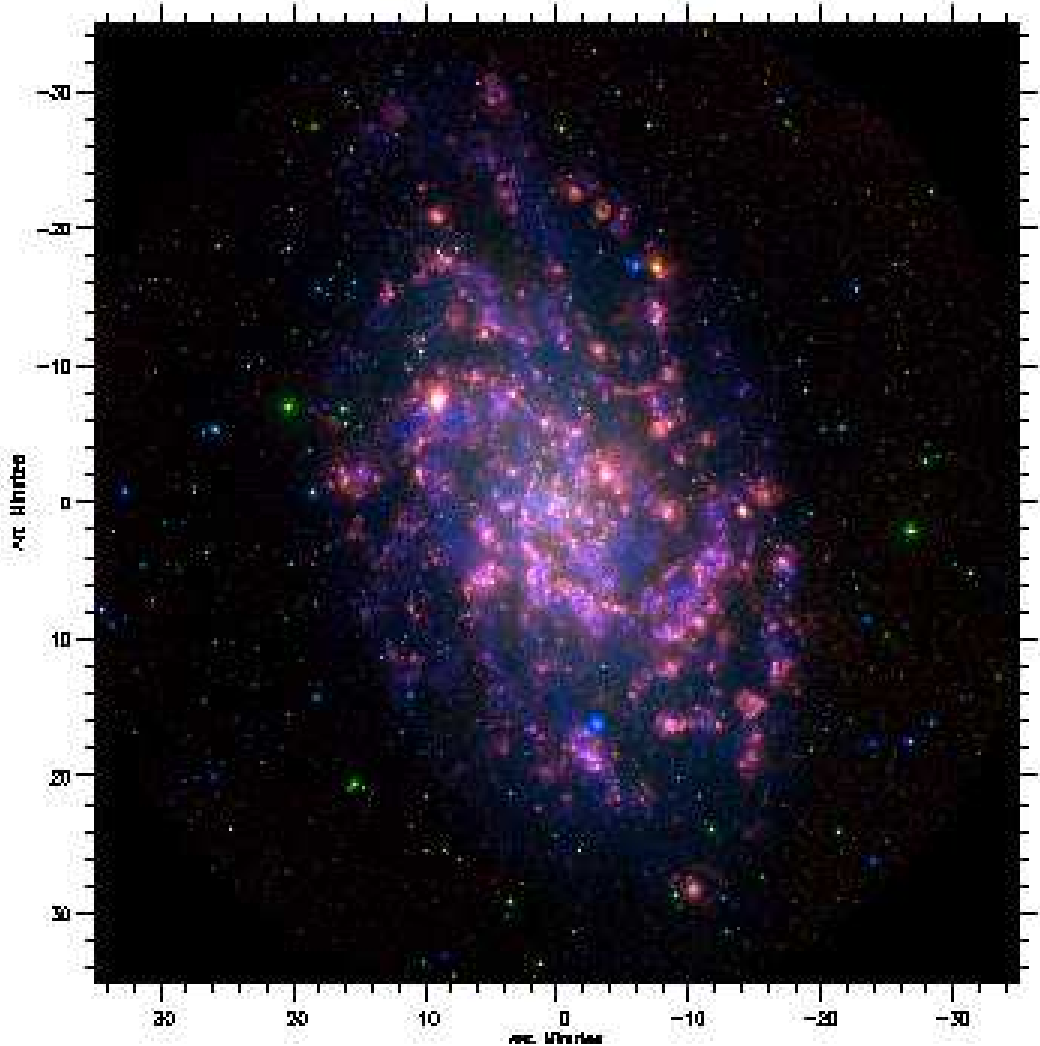}
\caption{Multi-wavelength view of M33, showing the same field as
Fig. 2 (11.4 kpc in projected extent).  The color channels of this
image are assigned as follows: $\Ha+{\rm continuum}$ (red), continuum
(green), GALEX NUV (blue). Differences in overall NUV and $\Ha$
morphology highlight the advantage of UV observations as a direct
tracer of sites which have recently hosted star-formation, rather than
the environmentally dependent, indirect, and highly age-sensitive view
given by $\Ha$.  This is not meant to downplay the importance of $\Ha$
observations, but rather to highlight the importance of both datasets
whenever the goal is to obtain a comprehensive view of star-formation
over the past few hundred Myr.}
\end{figure}

\begin{figure}
\epsscale{0.85}
\plotone{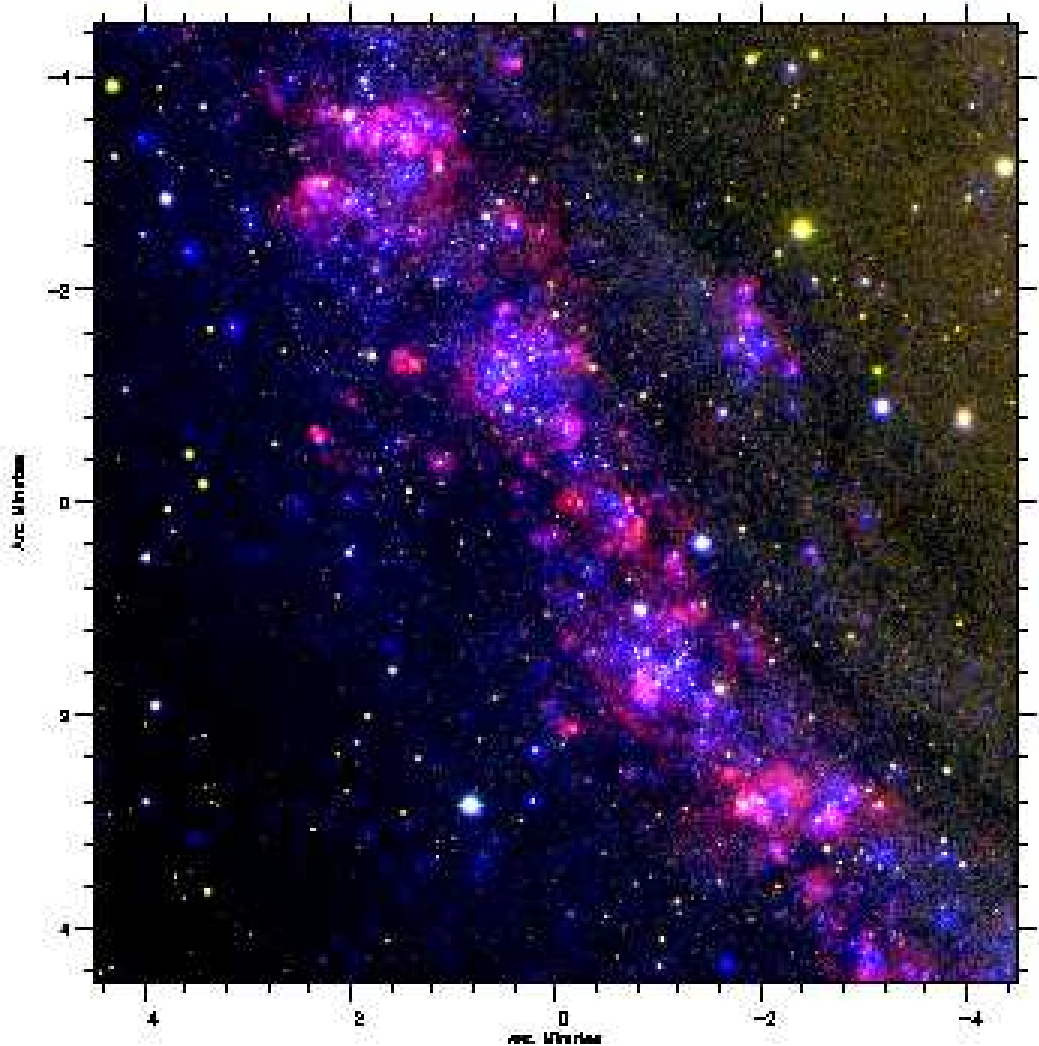}
\caption{A small section of M31's spiral arm structure, located to the
NE of the galaxy center.  The color channels of this image are
assigned as follows: $\Ha+{\rm continuum}$ (red), continuum (green),
GALEX NUV (blue). The detected distribution of NUV sources is clearly
more extensive than observed in $\Ha$, whereas the morphology of
emission line structures can provide valuable information on the
radiative- and mechanically-driven ISM feedback processes associated
with massive stars.}
\end{figure}
\end{document}